\documentclass[journal]{IEEEtran}
\usepackage{graphicx} 
\usepackage{booktabs}
\usepackage[table,xcdraw]{xcolor}
\usepackage{url}
\usepackage{tabularx}
\usepackage{booktabs}
\usepackage{adjustbox}
\usepackage{float}
\usepackage{amsmath}
\usepackage{cite}
\usepackage{subcaption}
 \usepackage{multirow}
\usepackage{graphicx}
\usepackage{algorithm}
\usepackage{algpseudocode}
\usepackage{amsfonts}

\ifCLASSINFOpdf

\else

\fi

\begin{document}

% \title{XAI-Driven Fraud Detection: An Attention-Based Ensemble of CNNs and Graph Neural Network for Credit Card Transactions}
\title{Explainable AI for Fraud Detection: An Attention-Based Ensemble of CNNs, GNNs, and A Confidence-Driven Gating Mechanism}

\author{Mehdi Hosseini Chagahi, Niloufar Delfan, Saeed Mohammadi Dashtaki, Behzad Moshiri (Senior Member, IEEE), Md. Jalil Piran (Senior Member, IEEE)

\thanks{(Corresponding Authors: B. Moshiri and M. J. Piran)}
\thanks{M. H. Chagahi, N. Delfan, and S. M. Dashtaki are with the School of Electrical and Computer Engineering, College of Engineering, University of Tehran, Tehran, Iran, (e-mail: mhdi.hoseini@ut.ac.ir;
niloufar.delfan@ut.ac.ir;
saeedmohammadi.d@ut.ac.ir
)}

\thanks{B. Moshiri is with the School of Electrical and Computer Engineering, College of Engineering, University of Tehran, Tehran, Iran and the Department of Electrical and Computer Engineering University of Waterloo,
Waterloo, Canada, (e-mail: moshiri@ut.ac.ir)}

\thanks{M. J. Piran is with the Department of Computer Science and Engineering, Sejong University, Seoul 05006, South Korea, (e-mail: piran@sejong.ac.kr)}}

\maketitle

\begin{abstract}
The rapid expansion of e-commerce and the widespread use of credit cards in online purchases and financial transactions have significantly heightened the importance of promptly and accurately detecting credit card fraud (CCF). Not only do fraudulent activities in financial transactions lead to substantial monetary losses for banks and financial institutions, but they also undermine user trust in digital services. This study presents a new stacking-based approach for CCF detection by adding two extra layers to the usual classification process: an attention layer and a confidence-based combination layer. In the attention layer, we combine soft outputs from a convolutional neural network (CNN) and a recurrent neural network (RNN) using the dependent ordered weighted averaging (DOWA) operator, and from a graph neural network (GNN) and a long short-term memory (LSTM) network using the induced ordered weighted averaging (IOWA) operator. These weighted outputs capture different predictive signals, increasing the model’s accuracy. Next, in the confidence-based layer, we select whichever aggregate (DOWA or IOWA) shows lower uncertainty to feed into a meta-learner.
To make the model more explainable, we use shapley additive explanations (SHAP) to identify the top ten most important features for distinguishing between fraud and normal transactions. These features are then used in our attention-based model. Experiments on three datasets show that our method achieves high accuracy and robust generalization, making it effective for CCF detection.
\end{abstract}


\IEEEpeerreviewmaketitle

\section{Introduction}
%\subsection{Background}
The convenience of financial services for consumers around the world has been tremendously improved by digital technology. However, this development also opened up new avenues for financial fraud, especially in credit card transactions \cite{mienye2024deep,lebichot2024assessment}.
% \cite{yang2021impact, yuan2021multigranulation,singh2022credit}
Looking at the numbers, one can tell that credit card fraud (CCF) is a big concern. According to Nilson's report; global losses from card fraud amounted to 28.65 billion dollars in 2019. These figures are expected surpassing 38.5 billion dollars by 2027 \cite{cherif2023credit,marchioni2023anomaly}. In the United States alone, CCF cases reported to the federal trade commission (FTC) were over 393 thousand during 2020, indicating a significant rise from the previous year \cite{madhurya2022exploratory,xie2022time}. The advent of internet-based transactions worsened matters leading into rise of fraud cases involving card-not-present (CNP) transactions, which are quite common currently \cite{mienye2024deep,gurun2018trust}.

Traditional fraud detection methods, such as rule-based systems and manual transaction reviews, are often inadequate in the face of evolving fraud patterns \cite{shibata2022digitalization,barmo2024analysis}. These approaches are limited in their ability to adapt to new tactics employed by fraudsters and frequently generate high false-positive rates, leading to unnecessary inconvenience for legitimate customers \cite{wang2021deep, li2020deep}. In contrast, machine learning (ML) and deep learning (DL) methodologies offer data-driven solutions that are capable of learning complex patterns and adapting to emerging fraud strategies \cite{khalid2024enhancing, han2022competition}. By analyzing vast amounts of transaction data, these techniques provide more accurate, scalable, and efficient ways to identify fraudulent transactions.

Recent progress in data-driven computational intelligence has introduced powerful classification algorithms, such as decision trees, random forests, gradient boosting, convolutional neural networks (CNNs), and recurrent neural networks (RNNs), specifically tailored for fraud detection \cite{zhao2022financial, ghosh2022spatio, valavan2023predictive, chen2022credit}. Additionally, ensemble techniques that synthesize the outputs of multiple models have shown notable success by exploiting the strengths of diverse classifiers \cite{zhao2022financial,kowsalya2024credit}. However, despite these achievements, critical challenges persist in balancing predictive accuracy ($A_c$) with interpretability, particularly within financial institutions that require clear, trustworthy insights into automated decisions \cite{chatterjee2024digital}.

Despite these notable advances, significant research gaps persist in CCF detection. First, many approaches still rely on a single classification model or a simple ensemble of models, which may fail to leverage complementary predictive signals and overlook more sophisticated aggregation strategies \cite{abd2023efficient}. Second, although graph neural networks (GNNs), CNNs, and recurrent architectures like RNNs and long short-term memories (LSTMs) each show promise individually, their combined outputs are often integrated through naive averaging or straightforward stacking, diluting critical predictive cues. Third, most current work does not adequately address model uncertainty, potentially leading to unreliable predictions, particularly in highly imbalanced datasets. Finally, interpretability remains a pressing concern: despite its importance in promoting stakeholder trust, the majority of methods do not offer transparent or user-friendly ways to understand feature importance and decision rationale. These limitations underscore the need for a more robust, uncertainty-aware, and interpretable framework that intelligently fuses multiple neural models to enhance both $A_c$ and trust in credit card fraud detection systems. Our main contributions to this study are outlined below:

\begin{itemize}
\item A diversity-based classifier selection is proposed, where an attention-based stacking framework is designed to leverage four distinct classifiers- CNN, RNN, LSTM, and GNN- to achieve a balanced trade-off between reliability and robustness. Specifically, CNN and RNN outputs are combined using the dependent ordered weighted averaging (DOWA) operator, while LSTM and GNN outputs are fused via the induced ordered weighted averaging (IOWA) operator.

\item A confidence-aware combination layer is introduced, enabling dynamic selection of the more reliable aggregate (DOWA or IOWA) based on uncertainty estimates. This best-performing output is then fed into a meta-learner, ensuring higher $A_c$, particularly under highly imbalanced conditions.

\item Interpretability is enhanced through the use of Shapley Additive Explanations (SHAP), allowing the identification of the top ten most influential features affecting model decisions. This feature-level insight improves transparency and facilitates stakeholder trust in the fraud detection process.

\item An extensive evaluation and benchmarking are conducted using three benchmark datasets, where the proposed method is compared against individual classifiers (CNN, RNN, LSTM, GNN). The results demonstrate superior performance and robust generalization, validating the effectiveness of the proposed framework in credit card fraud detection.
\end{itemize}
The remainder of this paper is organized as follows.
Section \ref{Methodology} introduces the datasets used in our experiments and identifies the key features driving the proposed approach. It then discusses the aggregation operators (DOWA and IOWA) and concludes with a detailed exposition of the system’s architecture.
Section \ref{Results} presents the experimental results, offering a comprehensive analysis of the findings.
Section \ref{discuss} provides a broader discussion of these results, highlighting their implications and suggesting potential avenues for future research. Finally, Section \ref{Conclusion} provides the concluding remarks.

\section{Proposed Attention-based Ensemble System for CCF Detection}
\label{Methodology}
In this section, we detail the methodology adopted to develop and evaluate our stacking-based approach for CCF detection. We begin by introducing the three benchmark datasets, highlighting their characteristics and the rationale behind their selection. We then describe the process for identifying key features using SHAP, focusing on the top attributes that distinguish fraudulent from legitimate transactions. Next, we explain how the DOWA and IOWA operators are employed to fuse the outputs of our neural network classifiers, CNN, RNN, LSTM, and GNN, in an attention mechanism. Finally, we outline the overall system architecture, including the confidence-based combination layer, which helps select the most reliable ensemble output before passing it to a meta-learner, ensuring both high $A_c$ and robust generalization.

\subsection{Datasets}
\label{dataset}
For fraud detection, we use three datasets, each containing anonymized transaction-related features labeled $V1$ through $V28$ to preserve user privacy. These features capture various aspects of the transaction process, ranging from timing information to geographical context. Each dataset also includes the feature Amount, representing the monetary value of the transaction, and the binary attribute Class, where transactions are labeled as fraudulent (1) or legitimate (0). Table \ref{datasets} provides a statistical overview of all three datasets. Notably, Dataset 3 was constructed by concatenating Dataset 1 and Dataset 2, resulting in a larger sample size for more comprehensive analysis and model training.

To rigorously assess our framework in both balanced and imbalanced scenarios, we first employ a balanced dataset (Dataset 1) for initial training and testing. This approach allows us to observe the model’s baseline performance when both classes are represented more evenly. Next, we train our proposed framework on the highly imbalanced Dataset 2 and subsequently evaluate it on this same dataset to investigate how the skew in class distribution influences predictive $A_c$. Finally, we tune the model’s hyperparameters on the balanced dataset and test the refined model on the unbalanced dataset, mirroring real-world conditions where fraudulent transactions constitute only a small fraction of all transactions. Since the features in two datasets were anonymized, we assumed their equivalence when training on Dataset 1 and testing on Dataset 2. This step-by-step procedure ensures that our model is both reliable under balanced conditions and robust to the challenges posed by heavily imbalanced data.

\begin{table}[t]
    \centering
    \caption{Data partition.}
    \label{datasets}
    \begin{tabular}{llrrr}
        \toprule
        \textbf{Datasets} & & \textbf{samples} & \textbf{Fraud} & \textbf{Imbalance ratio} \\
        \midrule
    \multirow{2}{*}{Public 1} & Train set & 454 905 & 227 452 & 50\% \\
        & Test set & 113 725 & 56 863 & 50\% \\
        \midrule
        \multirow{2}{*}{Public 2} & Train set &  227 845 & 394 & 0.17\% \\
        & Test set & 56 962 &  98 & 0.17\% \\
        \midrule
        \multirow{2}{*}{Public 3} & Train set & 568 630 & 284 315 & 50\% \\
        & Test set & 284 807 &  492 & 0.17\% \\

        \bottomrule
    \end{tabular}
\end{table}

\begin{figure}[t]
    \centering
    \resizebox{9cm}{7cm}{\includegraphics{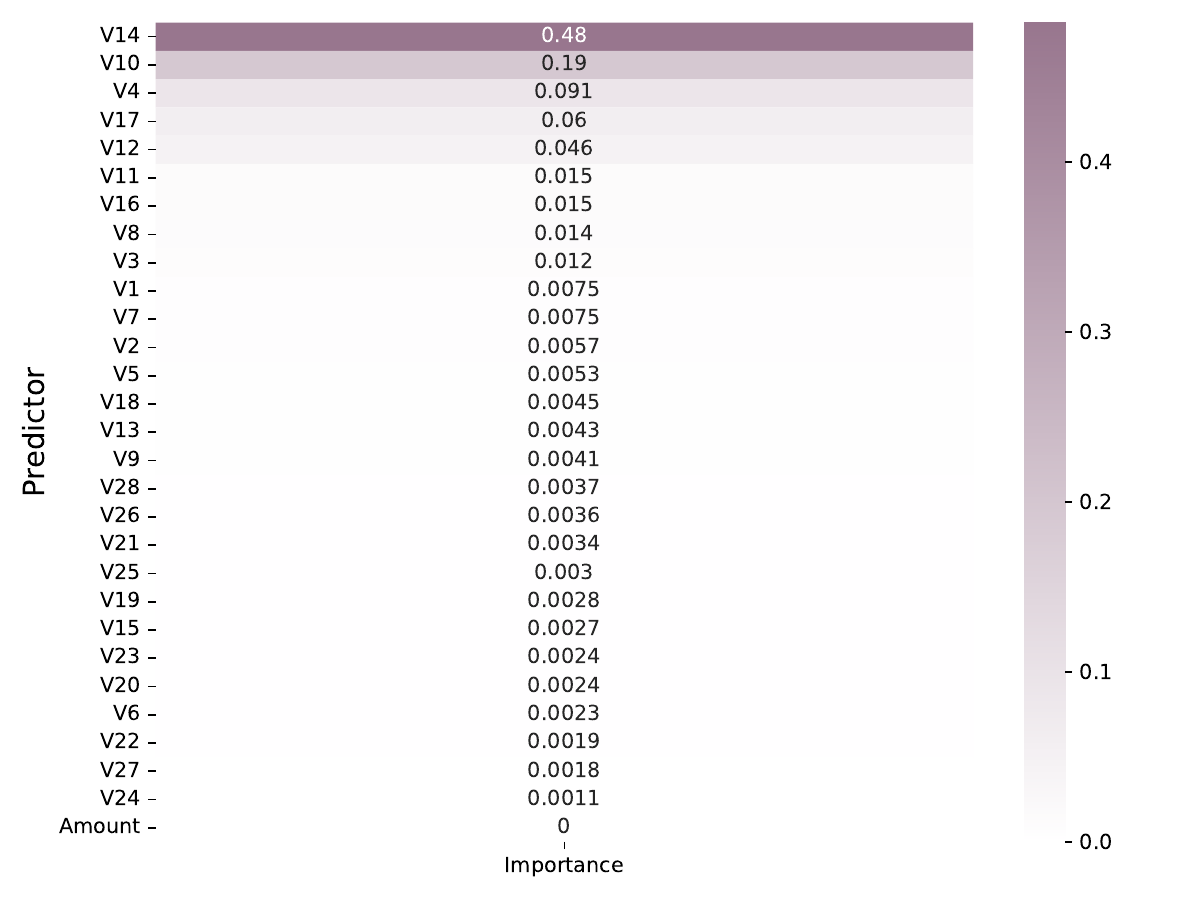}}
    \caption{SHAP-based feature importance ranking for fraud detection.}
    \label{shap}
\end{figure}

\subsection{Feature Importance Analysis Using SHAP}
\label{SHAP}
We employ SHAP with a bootstrap-based approach using 100 decision trees to assess feature importance in our fraud detection model. The results, visualized in Fig. \ref{shap}, rank features by their average SHAP values.
The analysis highlights $V14$, $V10$, and $V4$ as the most influential features, while others, such as $V24$ and $V27$, have minimal impact. Interestingly, the $Amount$ feature contributes negligibly, indicating that fraud detection relies more on transaction behavior than monetary values.
To improve efficiency and reduce complexity, we selected only the top 10 features ($V14$, $V10$, $V4$, $V17$, $V12$, $V11$, $V16$, $V8$, $V3$, and $V1$) for the final model. This selection enhances interpretability, reduces overfitting risks, and ensures practical deployment while maintaining high predictive performance.

\subsection{Aggregation Operators, DOWA and IOWA}
The Ordered Weighted Averaging (OWA) operator is a mathematical tool commonly employed in decision-making and aggregation tasks. It arranges inputs in a specified order before applying designated weights, as highlighted in \cite{chagahi2024cardiovascular}. In this study, Definition 1 introduces the DOWA operator, while Definition 2 presents the IOWA operator. In section \ref{proposed system}, each operator is used independently to merge the predictions of the first-layer classifiers within our proposed framework.

\textbf{Definition 1.}
Let $p_1, p_2, ..., p_n$ represent the probabilistic outputs or forecasts from initial classifiers regarding a particular sample, indicating the likelihood of the sample’s classification into each category. Let $m$ denote the mean of these predictions across all classes, i.e., $m = \frac{1}{n}\displaystyle\sum\limits_{j=1}^n p_j, \ (\alpha(1), \alpha(2),...,\alpha(n))$ is a permutation of $(1,2,...,n)$ such that $p_{\alpha(j-1)} \geq p_{\alpha(j)}$ for all $j = 2,...,n$, then we call
\begin{equation}
    c(p_{\alpha(j)},m)= 1-\frac{|p_{\alpha(j)}-m|}{\sum\limits_{j=1}^n |p_j-m|} , \ \ \ \ j = 1,2,...,n,  \label{q2}
\end{equation}
the similarity degree between the $j$-th highest value $p_{\alpha(j)}$ and the mean value $m$.

The DOWA operator calculates the weighted mean of the predictions using a set of weights, each of which emphasizes the relevance or significance of the respective prediction. Denote by $w = (w_1, w_2, ..., w_n)^T$ the weight vector for the OWA operator, and we establish: 

\begin{equation}
    w_j = \frac{c(p_{\alpha(j)},m)}{\sum\limits_{j=1}^n c(p_{\alpha(j)},m)} , \ \ \ \ j = 1,2,...,n, \label{q3}
\end{equation}
where $c(p_{\alpha(j)},m)$ is defined by (\ref{q2}). Clearly, we have $w_j \in [0,1]$ and $\sum\limits_{j=1}^n w_j = 1$. Since 
\begin{equation}
    \sum\limits_{j=1}^n c(p_{\alpha(j)},m) = \sum\limits_{j=1}^n c(p_{j},m),\label{q4}
\end{equation} 
then (\ref{q3}) can be reformulated as 
\begin{equation}
    w_j = \frac{c(p_{\alpha(j)},m)}{\sum\limits_{j=1}^n  c(p_{j},m)} , \ \ \ \ j = 1,2,...,n, \label{w_DOWA_Eq}
\end{equation} 
under these circumstances, it follows that 
{\begin{equation}  
    OWA (p_1,p_2,...,p_n) = \frac{ \sum\limits_{j=1}^n c(p_{\alpha(j)},m)p_{\alpha(j)}}{\displaystyle  \sum\limits_{j=1}^n c(p_{j},m)}. \label{DOWA_Eq}
\end{equation}}

Using (\ref{w_DOWA_Eq}) weights corresponding to the predictions of each classifier are calculated. These weights can have different values for each sample. Then, the predictions of the first layer classifiers are aggregated based on the weights assigned to them by (\ref{DOWA_Eq}). This operator effectively converts the separate and independent predictions of single first layer classifiers into a single prediction.

\textbf{Definition 2.}
The development of the weighting vector is a critical aspect widely discussed, as seen in \cite{yager2016some,dashtaki2022stock}. Filev and Yager proposed a method in \cite{filev1998issue} for deriving the weighting vector for an OWA aggregation through the analysis of observational data. This approach closely aligns with learning algorithms typical of neural networks, as referenced in \cite{khaled2023dowg,wang2022study}, utilizing the gradient descent method as its foundation. This section outlines the process for determining the OWA weighting vector’s weights based on observational input.

We are presented with a set of samples, each comprising a series of values $(p_{k1},p_{k2},...,p_{kn})$, referred to as the arguments, and a corresponding single value named the aggregated value, denoted as $d_k$. Our aim is to develop a model that accurately captures this aggregation process using the OWA method. In essence, the task is to derive a weighting vector $W$ that effectively represents the aggregation mechanism across the dataset. The aim is to determine the weighting vector $W$ based on the data provided.

The task focuses on simplifying the challenge of identifying the weights by leveraging the OWA aggregation’s linear nature with the reordered inputs. For the $k$th data point, the reordered arguments are signified as $b_{k1}, b_{k2}, ..., b_{kn}$, with $b_{kj}$ being the $j$th largest value within the set $(p_{k1}, p_{k2}, ..., p_{kn})$. With the arguments now ordered, the objective turns into determining the OWA weight vector $W^{T} = [w_1, w_2, ..., w_n]$ in such a manner that the equation $b_{k1}w_1 + b_{k2}w_2 + ... + b_{kn}w_n = d_k$ holds true for each instance $k$, from $1$ up to $N$.

We will adjust our method by searching for a weights vector, $W$, for OWA, aiming to closely resemble the aggregating function by reducing immediate errors $e_k$
\begin{equation}
 e_k = \frac{1}{2}\ (b_{k1}w_1 +b_{k2}w_2+...+b_{kn}w_n-d_k)^{2}.   
 \label{eq 7}
\end{equation}

In relation to the weights $w_i$, this learning challenge represents a constrained optimization issue. This is because the OWA weights $w_i$ must adhere to two specific criteria: $\sum\limits_{i=1}^n w_i = 1$ and  $w_i \in [0,1]$ for $i = 1$ to $n$.

To bypass the restrictions, each of the OWA weights is represented in the following manner.
\begin{equation}
  w_i = \frac{e^{\beta{i}}}{\sum\limits_{j=1}^n e^{\beta{j}}}.
  \label{eq 8}
\end{equation}

With the transformation, regardless of the parameter values $\beta_i$, the weights $w_i$ will fall within the unit interval and their sum will equal 1. Hence, the original constrained minimization issue is converted into an unconstrained nonlinear programming problem:

\textbf{Minimize the immediate errors } $\mathbf{e_k}$ \textbf{:}

By integrating (\ref{eq 8}) into (\ref{eq 7}), it can be stated:

\begin{equation}
\resizebox{\linewidth}{!}{$
    e_{k} = \frac{1}{2}\ (b_{k1}\frac{e^{\beta_{1}}}{\sum\limits_{j=1}^n e^{\beta_{j}}}+b_{k2}\frac{e^{\beta_{2}}}{\sum\limits_{j=1}^n e^{\beta_{j}}}+...+b_{kn}\frac{e^{\beta_{n}}}{\sum\limits_{j=1}^n e^{\beta_{j}}}-d_{k})^{2}$}
\end{equation}

Motivated by the significant achievements of gradient descent methodologies in the backpropagation approach utilized for training in neural networks, we adopt it here. By applying the gradient descent technique, we establish the subsequent rule for parameter updates $\beta_i , i =(1,n)$;
\begin{equation}
\beta_i(l+1) = \beta_i (l)-\alpha\frac{\partial e_k}{\partial \beta_i}|_{\beta_i = \beta_i(l)},
\end{equation}
where $\alpha$ represents the rate of learning $(0<\alpha<1)$ and $\beta_i (l)$ signifies the approximation of $\beta_i$ following the $l$th iteration.

To simplify notation, we represent the estimate of the aggregated value $d_k$ as $\hat{d}_k$
\begin{equation}
\hat{d}_k = b_{k1}w_1+b_{k2}w_2+...+b_{kn}w_n.
\end{equation}
Then, regarding the partial derivative $\frac{\partial e_k}{\partial \beta_i}$ we get

\begin{equation}
\frac{\partial e_k}{\partial \beta_i} = w_i (b_{ki}-\hat{d}_k)(\hat{d}_k-d_k),\ \ \  i=[1,n].
\end{equation}
Finally, we establish the formula for adjusting the parameters $\beta_i$ in the following manner:
\begin{equation}
\beta_i(l+1) = \beta_i(l)-\alpha w_i (b_{ki}-\hat{d}_k)(\hat{d}_k-d_k),
\end{equation}
where the $w_i$ values are determined at each iterative step based on the current estimates of the $\beta_i$ values
\begin{equation}
 w_i = \frac{e^{\beta_{i}(l)}}{\sum\limits_{j=1}^n e^{\beta_{j}(l)}}, \ \ i=[1,n],
 \end{equation}
and $\hat{d}_k$ represents the current approximation of the aggregated values $d_k$.

This outlines the procedure implemented at each step of the iteration.
\begin{itemize}
    \item We possess a current estimation of the $\beta_i$, denoted as $\beta_i(l)$, and a new observation composed of the ordered arguments $b_{k1}, b_{k2}, ..., b_{kn}$ along with an aggregated value $d_k$.

    \item We utilize $\beta_i(l)$ to generate a current estimation for the weights
\begin{equation}
w_i = \frac{e^{\beta_{i}(l)}}{\sum\limits_{j=1}^n e^{\beta_{j}(l)}}.
\end{equation}
\item We apply the estimated weights and the ordered arguments to derive a computed aggregated value
\begin{equation}
\hat{d}_k = b_{k1}w_1(l)+b_{k2}w_2(l)+...+b_{kn}w_n(l).
\end{equation}
\item We update our estimations of the $\lambda_i$.
\begin{equation}
\beta_i(l+1) = \beta_i(l)-\alpha w_i(l) (b_{ki}-\hat{d}_k)(\hat{d}_k-d_k).
\end{equation}
\end{itemize}
In summary, both the DOWA and IOWA operators offer a systematic approach to aggregating probabilistic outputs by considering the order and relative similarity of predictions. By learning appropriate weights from training data and then applying them to sorted inputs, these operators capture subtle variations in classifier outputs, thereby enhancing the overall ensemble’s reliability. In the following section, we detail how these aggregation techniques are integrated into our proposed fraud detection framework, illustrating their role in the model’s multi-layer architecture.

\subsection{ CCF detection Framework: Structure, Development, and Examination}
\label{proposed system}

\begin{figure}[t]
\centering
\includegraphics[scale=0.5]{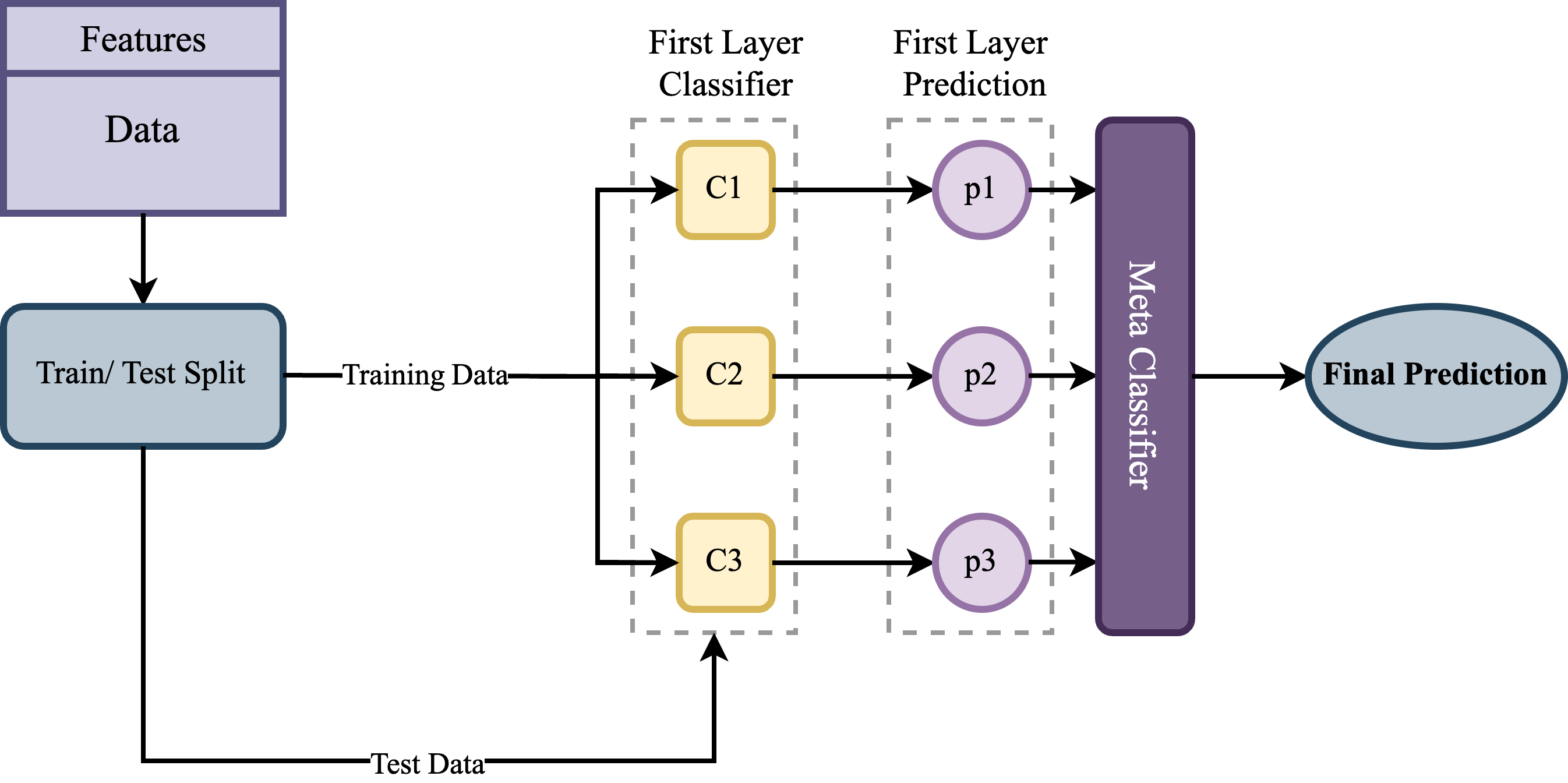}
\caption{Traditional stacking classifier}
\label{stacking}
\end{figure}

\begin{figure*}[t]
\centering
\includegraphics[scale=1]{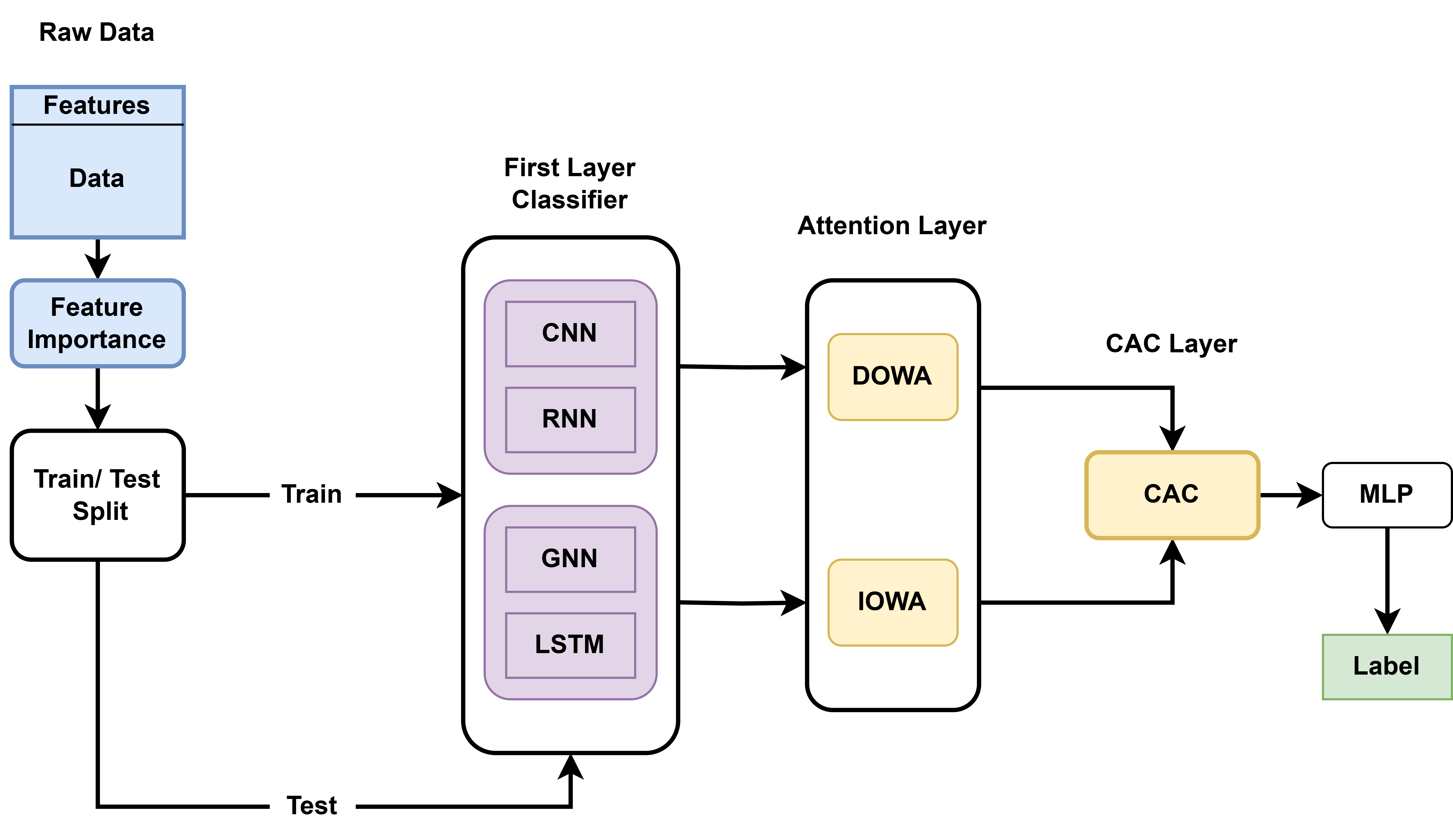}
\caption{The block diagram of proposed architecture for detecting CCF.}
\label{flow chart}
\end{figure*}

Fig. \ref{stacking} illustrates the structure of the traditional stacking classifier, an ensemble learning technique that improves predictive $A_c$ by combining multiple ML models. In this method, several learners are trained on the dataset, and their predictions are then used as input features for a meta-learner. The meta-learner learns how to optimally combine the outputs of the first layer models to make the final prediction.

This study enhances the traditional stacking architecture by incorporating an attention layer and a confidence-aware combination layer. Fig. \ref{flow chart} shows the block diagram of the proposed architecture for CCF detection. In the following, we delve into the details of this entirely attention-based ensemble architecture.

\subsubsection{Diversity-Based Selection Strategy} 
To evaluate the performance of deep learning models in fraud detection, we compare four architectures: CNN, RNN, LSTM, and GNN. These models are trained and optimized using cross-validation to ensure fair comparison. The main hyperparameters for each method are as follows.

 \textbf{(1) CNN} \cite{krizhevsky2017imagenet}: Number of convolutional layers = 3,
Kernel size = 3,
Stride = 1,
Pooling type = MaxPooling,
Pool size = 2,
Activation function = ReLU,
Fully connected layers = 2 (with 128 and 64 neurons),
Optimizer = Adam (learning rate = 0.001),
Epochs = 200.

 \textbf{(2) RNN} \cite{sherstinsky2020fundamentals}: Number of recurrent layers = 2,
Hidden units per layer = 128,
Activation function = Tanh,
Dropout rate = 0.2,
Optimizer = RMSprop (learning rate = 0.001),
Epochs = 200,

 \textbf{(3) LSTM} \cite{sherstinsky2020fundamentals}: Number of LSTM layers = 2,
Hidden units per layer = 128,
Activation function = Tanh,
Dropout rate = 0.3,
Recurrent dropout = 0.2,
Optimizer = Adam (learning rate = 0.0005),
Epochs = 200.

 \textbf{(4) GNN} \cite{zhou2020graph}: Number of graph convolution layers = 3,
Hidden units per layer = 64,
Graph aggregation method = Mean pooling,
Activation function = LeakyReLU,
Dropout rate = 0.2,
Optimizer = Adam (learning rate = 0.001),
Epochs = 200.

\begin{figure}[t]
    \centerline{\includegraphics[scale=0.7]{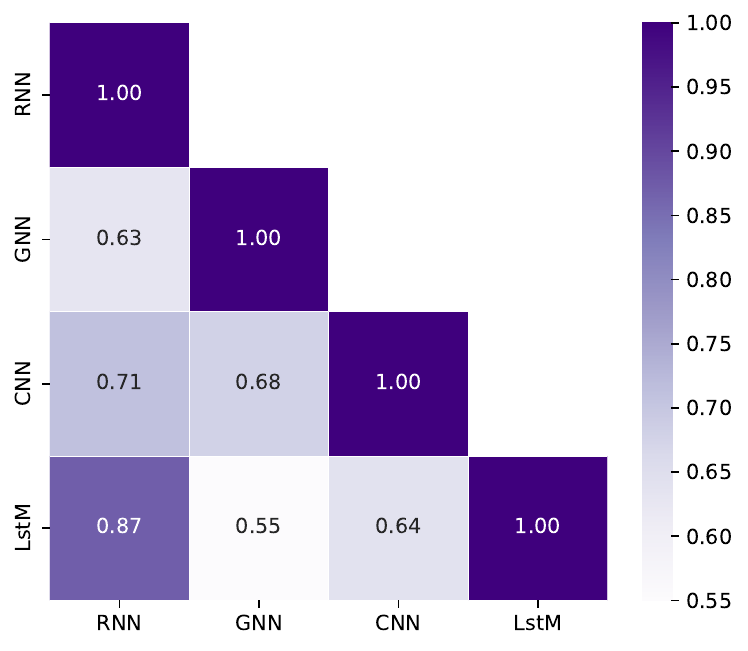}}
	\caption{Correlation matrix illustrating the diversity among predictions from RNN, GNN, CNN, and LSTM. Darker shades indicate higher agreement, whereas lighter shades reflect lower correlation.}
	\label{scatter}
\end{figure}

Based on the correlation matrix in Fig. \ref{scatter}, we observe that RNN and LSTM have the highest correlation (0.87), indicating considerable overlap in their predictive patterns. Consequently, combining them would offer limited diversity and might not significantly improve overall performance. By contrast, CNN and RNN (correlation: 0.71) exhibit moderate similarity, providing enough redundancy for reliability while maintaining enough difference for enhanced coverage. Similarly, GNN and LSTM show the lowest correlation (0.55), suggesting that each model captures distinct aspects of the data and could thus complement one another well. Therefore, in our framework, CNN and RNN are aggregated via the DOWA operator, while GNN and LSTM are combined using the IOWA operator. This diversity-based selection ensures a balanced trade-off between redundancy (stability) and complementarity (robustness), ultimately improving the ensemble’s ability to detect fraud across a variety of transaction patterns.

\subsubsection{Attention Layer: DOWA and IOWA Combinations}
In our proposed framework, we first introduce an attention layer that uses the DOWA operator to combine the soft outputs of CNN and RNN. DOWA assigns higher weights to classifiers whose predictions lie closer to the overall mean, preventing any single classifier with inconsistent outputs from excessively skewing the final aggregate. Since CNN and RNN exhibit a moderate correlation (0.71), this approach balances reliability and diversity by emphasizing their areas of agreement while retaining their distinct predictive signals, ultimately yielding a more robust and stable ensemble.

Building on this, we extend the attention layer by incorporating the IOWA operator to merge the soft outputs of LSTM and GNN. In contrast to DOWA, which highlights predictions near the mean, IOWA weights inputs according to their order and assigned significance, making it especially suitable for combining classifiers with low correlation. Given that LSTM and GNN demonstrate the lowest correlation (0.55), this weighted strategy exploits their complementary strengths, LSTM’s ability to capture temporal patterns and GNN’s capacity for graph-structured data representation. By carefully assigning weights through IOWA, our framework amplifies reliable signals from each model while minimizing the influence of outlier predictions, further strengthening the ensemble’s overall performance in fraud detection.

Therefore, the DOWA operator is used to compute weights for the RNN and CNN classifiers, and the IOWA operator does the same for the GNN and LSTM classifiers. Once these weights are learned, the soft probabilities from each first-layer classifier are combined accordingly and sorted in descending order. For a sample $i \in \{1,2,\dots,N\}$, the DOWA operator produces a two-dimensional output of the form  
$
F_{\mathrm{DOWA}} = \bigl[f\bigl(p_{\mathrm{RNN}}^0, p_{\mathrm{CNN}}^0\bigr),\, f\bigl(p_{\mathrm{RNN}}^1, p_{\mathrm{CNN}}^1\bigr)\bigr]^T,
$
while the IOWA operator yields  

$F_{\mathrm{IOWA}} = \bigl[f\bigl(p_{\mathrm{GNN}}^0, p_{\mathrm{LSTM}}^0\bigr),\, f\bigl(p_{\mathrm{GNN}}^1, p_{\mathrm{LSTM}}^1\bigr)\bigr]^T$.
 
Here, $p_{\mathrm{RNN}}^0$ denotes the probability that sample \(i\) belongs to class 0 as predicted by the RNN, and $f\bigl(p_{\mathrm{RNN}}^0, p_{\mathrm{CNN}}^0\bigr)$ represents the fused value of the respective class-0 probabilities from the RNN and CNN. Because each of these operators aggregates the outputs of two classifiers, they effectively map from $\mathbb{R}^4$ (the four class probabilities, two from each classifier) to a reduced space $\mathbb{R}^2$.
\subsubsection{Confidence-Aware Combination (CAC) Layer}
After generating two fused probability vectors, $F_{\mathrm{DOWA}}$ from the CNN–RNN pair and $F_{\mathrm{IOWA}}$ from the GNN–LSTM pair, the framework applies a confidence-based mechanism to decide which set of fused probabilities is passed to the meta-learner. Specifically, we quantify uncertainty by comparing the absolute difference between class-0 and class-1 probabilities for each aggregator. Formally, for each sample $i$,
$
\Delta_{\mathrm{DOWA}} = \Bigl\lvert f\bigl(p_{\mathrm{RNN}}^0,\, p_{\mathrm{CNN}}^0\bigr) \;-\; f\bigl(p_{\mathrm{RNN}}^1,\, p_{\mathrm{CNN}}^1\bigr)\Bigr\rvert
\quad\text{and}\quad
\Delta_{\mathrm{IOWA}} = \Bigl\lvert f\bigl(p_{\mathrm{GNN}}^0,\, p_{\mathrm{LSTM}}^0\bigr) \;-\; f\bigl(p_{\mathrm{GNN}}^1,\, p_{\mathrm{LSTM}}^1\bigr)\Bigr\rvert.
$

We interpret the operator with the smaller absolute difference as having higher uncertainty. Consequently, if $\Delta_{\mathrm{IOWA}}\ \le \Delta_{\mathrm{DOWA}}$, we feed $F_{\mathrm{DOWA}}$ (the DOWA-fused output) into a multi-layer perceptron (MLP) meta-learner; otherwise, we select $F_{\mathrm{IOWA}}$. As detailed in Algorithm 1, this decision process is repeated for each sample. By always routing the more confident fused probabilities to the meta-learner, the final classification performance is enhanced, particularly under challenging conditions such as imbalanced datasets.

\begin{algorithm}[t]
\caption{: Confidence-Aware aggregator selection
choosing between the DOWA- and IOWA-fused outputs based on uncertainty measures for each sample.}
\label{algo}
\begin{algorithmic}
\State \textbf{Input:}  
$f_{i}(p_{RNN}^0, p_{CNN}^0)$, $f_{i}(p_{RNN}^1, p_{CNN}^1)$, $f_{i}( p_{GNN}^0, p_{LSTM}^0)$, $f_{i}( p_{GNN}^1, p_{LSTM}^1)$
\State \textbf{Output:} One of the aggregate values, $F_{DOWA}$ or $F_{IOWA}$, for each sample
\State $i \gets 1$
\While{$i \leq N$}
\If{$|f_{i}(p_{RNN}^0, p_{CNN}^0) - f_{i}(p_{RNN}^1, p_{CNN}^1)| \geq |f_{i}( p_{GNN}^0, p_{LSTM}^0) - f_{i}( p_{GNN}^1, p_{LSTM}^1)|$}
    \State \textbf{return} $F_{DOWA}$
\Else
    \State \textbf{return} $F_{IOWA}$
\EndIf
\State $i \gets i + 1$
\EndWhile
\end{algorithmic}
\end{algorithm}

\subsubsection{Meta-Learner Configuration } 
In our final classification stage, we employ a MLP as the meta-learner to process the fused probabilities (i.e., $F_{\mathrm{DOWA}}$ or $F_{\mathrm{IOWA}}$ selected by the confidence-aware combination layer. The MLP architecture consists of two hidden layers, each containing $32$ neurons with ReLU activation, to capture non-linear relationships between the fused features. A dropout rate of $0.2$ is applied to mitigate overfitting and improve generalization. We train the MLP using the Adam optimizer with a learning rate of $10^{-3}$, a batch size of $64$, and early stopping based on a validation loss threshold. This configuration ensures both effective learning and a balance between model complexity and computational efficiency.

\begin{figure}[t]
    \centering
    \resizebox{9cm}{7cm}{\includegraphics{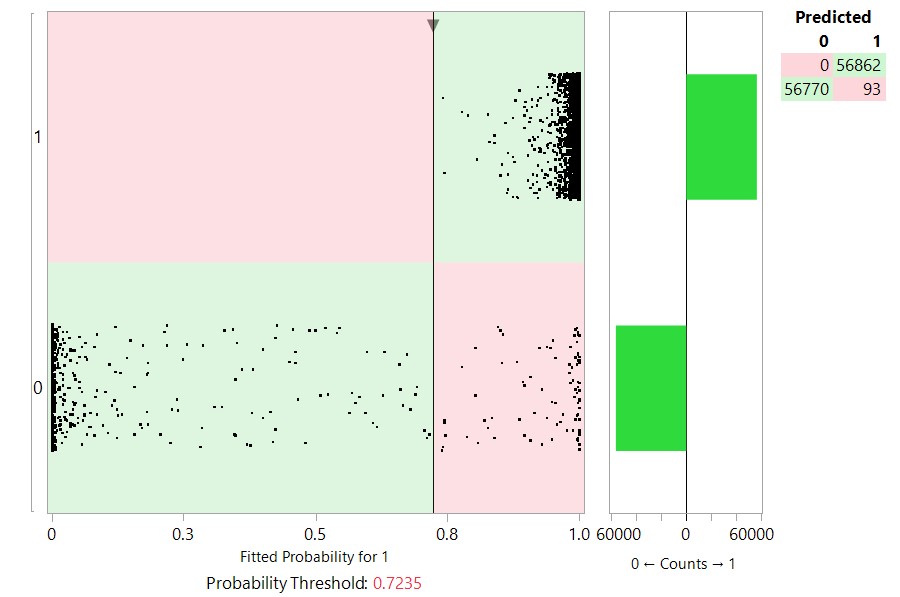}}
    \caption{Confusion matrix for Dataset 1, using the proposed ensemble model with an 80-20 stratified train-test split.}
    \label{conf1}
\end{figure}

\begin{figure}[t]
    \centering
    \resizebox{9cm}{6cm}{\includegraphics{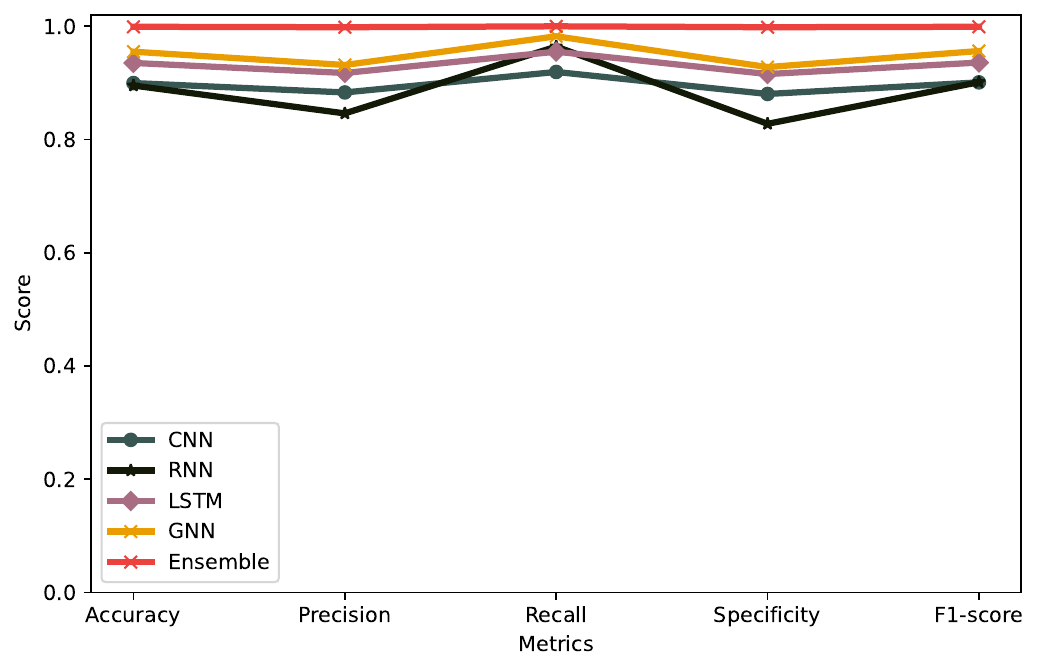}}
    \caption{Performance comparison of CNN, RNN, LSTM, GNN, and the proposed ensemble model on Dataset 1.}
    \label{data1}
\end{figure}

\begin{figure}[t]
    \centering
    \resizebox{9cm}{7cm}{\includegraphics{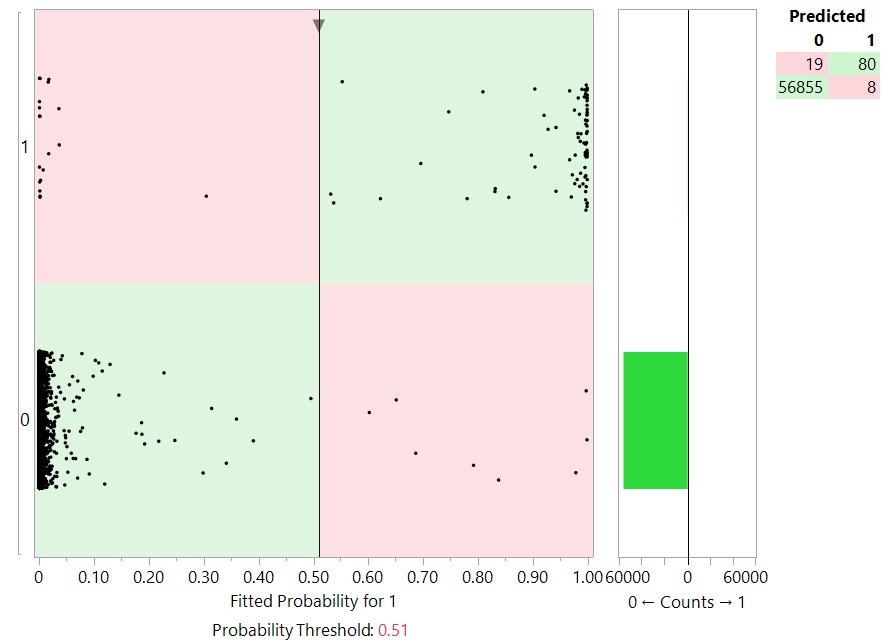}}
    \caption{Confusion matrix for Dataset 2, using the proposed ensemble model with an 80-20 stratified train-test split.}
    \label{conf2}
\end{figure}

\begin{figure}[t]
    \centering
    \resizebox{9cm}{6cm}{\includegraphics{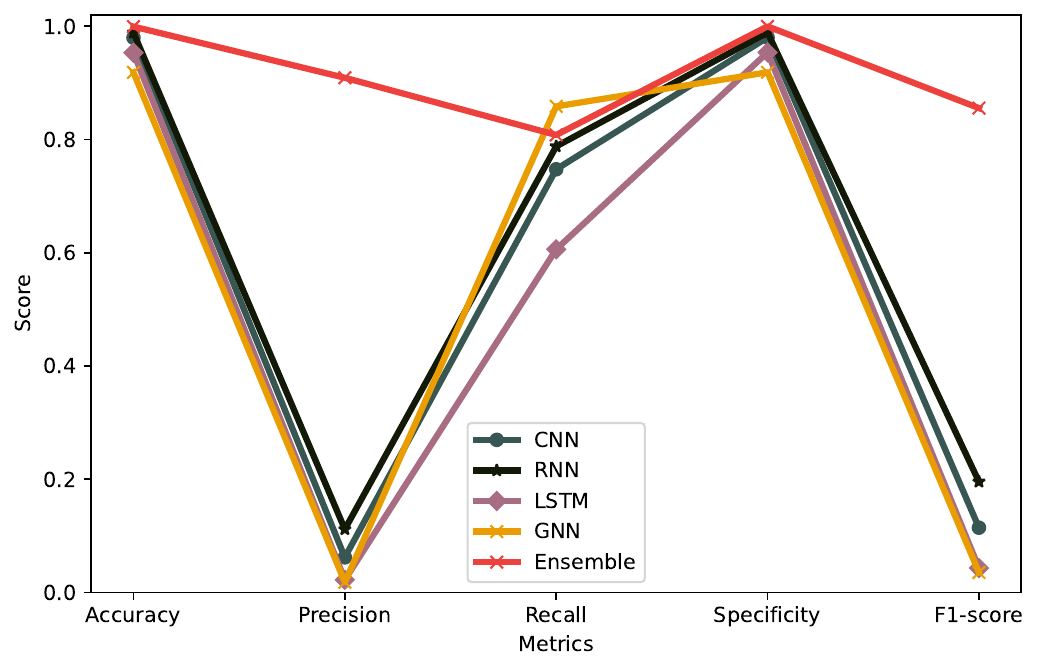}}
    \caption{Performance comparison of CNN, RNN, LSTM, GNN, and the proposed ensemble model on Dataset 2.}
    \label{data2}
\end{figure}

\begin{figure}[t]
    \centering
    \resizebox{9cm}{7cm}{\includegraphics{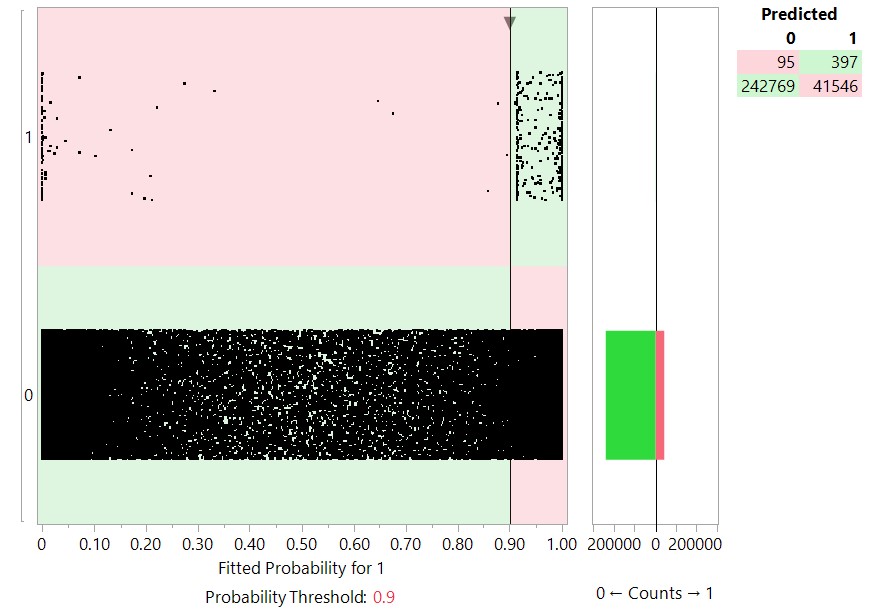}}
    \caption{Confusion matrix for Dataset 3, where the ensemble model was trained on balanced Dataset 1 and tested on imbalanced Dataset 2.}
    \label{conf3}
\end{figure}

\begin{figure}[t]
    \centering
    \resizebox{9cm}{6cm}{\includegraphics{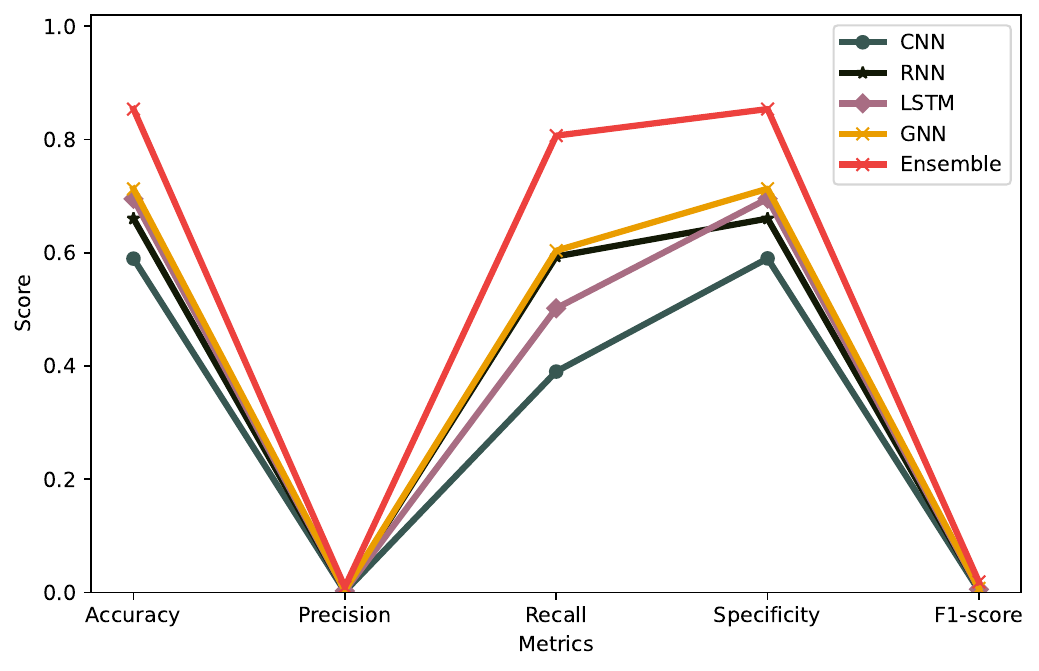}}
    \caption{Performance comparison of CNN, RNN, LSTM, GNN, and the proposed ensemble model on Dataset 3.}
    \label{data3}
\end{figure}

\begin{figure}[t]
    \centering
    \resizebox{9cm}{6cm}{\includegraphics{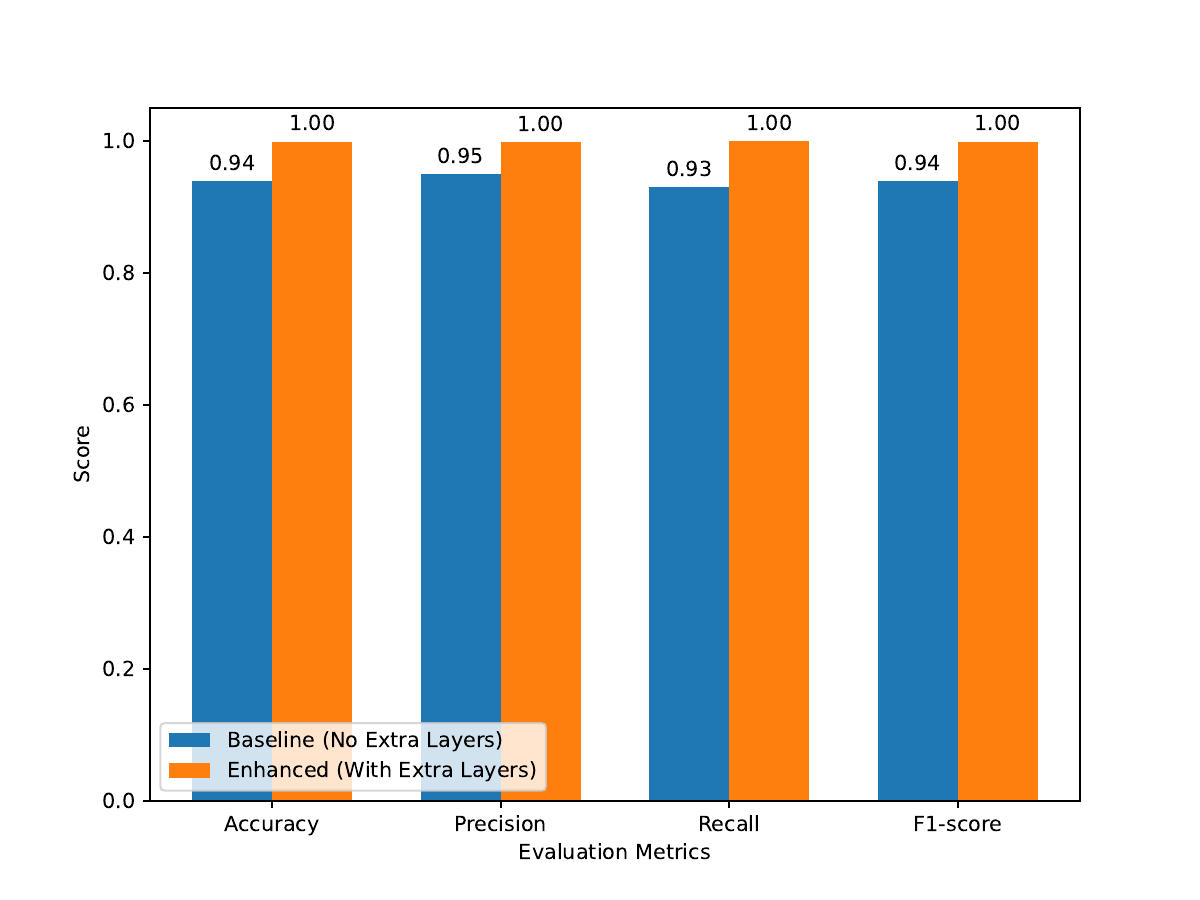}}
    \caption{Impact of incorporating the attention and confidence-aware combination layers in the stacking model on the test data of Dataset 1.}
    \label{data1_layer}
\end{figure}

\begin{figure}[t]
    \centering
    \resizebox{9cm}{6cm}{\includegraphics{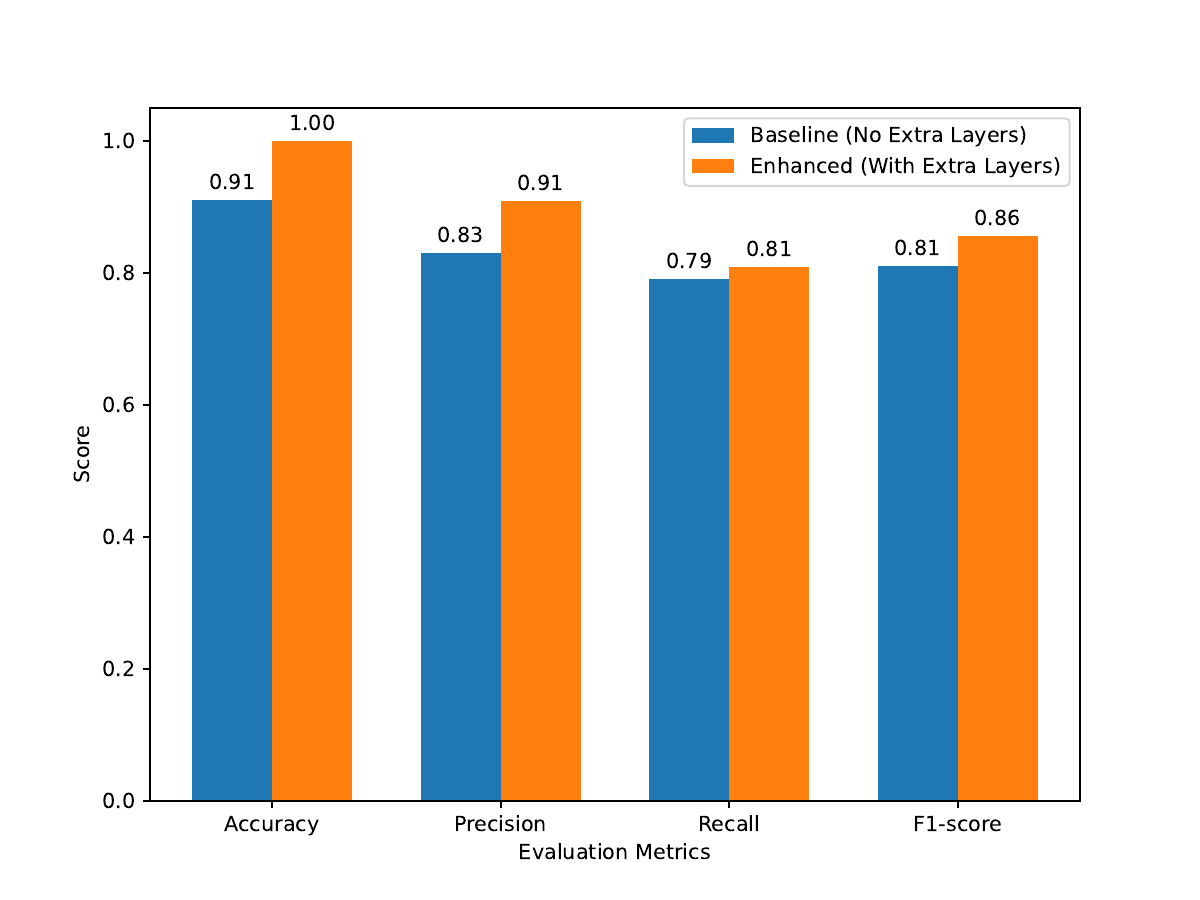}}
    \caption{Impact of incorporating the attention and confidence-aware combination layers in the stacking model on the test data of Dataset 2.}
    \label{data2_layer}
\end{figure}

\section{Results}
\label{Results} 
In this section, we delve deeper into the results of our proposed algorithm and present a comprehensive set of performance metrics, including $A_c$, recall ($R_e$), precision ($P_r$), specificity ($S_p$), and the F1-Score ($F_1$). The subsequent figures illustrate how the attention-based ensemble model performs on three different datasets. For each dataset, we provide two visualizations: a confusion matrix, which displays classification outcomes for the test set, green areas indicate correct predictions and red areas denote misclassifications, and a performance comparison graph, contrasting the ensemble’s key evaluation metrics against those of the individual deep learning models (CNN, RNN, LSTM, and GNN). The fitted probability threshold is shown in each confusion matrix, defining the boundary for classifying instances as either fraudulent (1) or legitimate (0). Additionally, a bar chart adjacent to each matrix offers a quick snapshot of how many samples fall into each category. This side-by-side comparison underscores the benefits of our attention-based approach in terms of $A_c$, $R_e$, $P_r$, $S_p$, and the $F_1$.

Dataset 1 (balanced):  
In Fig. \ref{conf1} and Fig. \ref{data1}, the model is trained and tested with an 80–20 stratified split on a balanced dataset (Dataset 1). As shown in the confusion matrix (Fig. \ref{conf1}), the proposed model achieves a high true negative rate, effectively identifying legitimate transactions. However, the number of correctly identified fraud cases remains limited, an expected result given the overall lower prevalence of fraud in this dataset. The performance comparison (Fig. \ref{data1}) demonstrates that the ensemble model consistently surpasses individual deep learning models, with a particularly strong showing in $R_e$, a critical metric for fraud detection.

Dataset 2 (heavily imbalanced):  
For Dataset 2 (Fig. \ref{conf2} and Fig. \ref{data2}), the same 80–20 stratified split is applied, but the task becomes more challenging due to a significantly skewed class distribution. The confusion matrix (Fig. \ref{conf2}) reveals an increased number of false negatives, indicating that detecting fraudulent transactions is harder under these conditions. This likely stems from distinct fraud patterns and their lower representation in Dataset 2. Nonetheless, the ensemble model still outperforms each individual model, as evident in Fig. \ref{data2}, maintaining higher $R_e$ and $S_p$, even under difficult, highly imbalanced circumstances.

Dataset 3 (cross-dataset evaluation): 
In Dataset 3 (Fig. \ref{conf3} and Fig. \ref{data3}), we adopt a different training strategy: Dataset 1 (balanced) is used for training, while Dataset 2 (imbalanced) is reserved for testing. The confusion matrix (Fig. \ref{conf3}) shows that, although the model tends to be conservative, leading to a higher false positive rate, it nonetheless captures a substantial number of fraudulent transactions. Fig. \ref{data3} confirms that, despite this trade-off, our ensemble model continues to outperform the standalone classifiers, demonstrating a strong balance between $R_e$ and $S_p$ when transitioning from balanced training data to highly imbalanced real-world scenarios.

Fig. \ref{data1_layer} and Fig. \ref{data2_layer} evaluate the impact of incorporating attention-based fusion and confidence-aware combination layers into the stacking model by comparing two versions: a baseline model without these layers and an enhanced version that integrates them. The results show that adding these extra layers significantly improves performance. This improvement indicates that intelligently combining model outputs and reducing uncertainty in final predictions enhances fraud detection effectiveness.

\begin{table*}[ht!]
    \centering
    \caption{Performance comparison of various ensemble learning models and the proposed attention-based stacking model for CCF detection across multiple evaluation metrics.}
    \label{ensemble_attention}
    \begin{tabular}{lccccccccc}
        \toprule
        \textbf{Datasets} & \textbf{ } & \textbf{LightGBM} & \textbf{RF} & \textbf{XGBoost} & \textbf{DF} & \textbf{AdaBoost} & \textbf{ET} & \textbf{BDT} & \textbf{Proposed} \\
        \midrule
        \multirow{3}{*}{Dataset 1} 
        & $P_r$ & 0.9313 & 0.8727 & 0.9280 & 0.9238 & 0.9281 & 0.8853 & 0.8815 & 0.9984 \\
        & $R_e$ & 0.8606 & 0.9172 & 0.8172 & 0.8592 & 0.8595 & 0.8931 & 0.9114 & 1.0000 \\
        & $F_1$ & 0.8945 & 0.8942 & 0.8693 & 0.8901 & 0.8923 & 0.8885 & 0.8959 & \textbf{0.9992} \\
        \midrule
        \multirow{3}{*}{Dataset 2} 
        & $P_r$ & 0.8847 & 0.8158 & 0.8814 & 0.8799 & 0.8875 & 0.7433 & 0.8375 & 0.9091 \\
        & $R_e$ & 0.7596 & 0.8204 & 0.7400 & 0.7298 & 0.7596 & 0.8827 & 0.8327 & 0.8081 \\
        & $F_1$ & 0.8172 & 0.8181 & 0.8042 & 0.7975 & 0.8162 & 0.8079 & 0.8347 & \textbf{0.8556} \\
        \midrule
        \multirow{3}{*}{Dataset 3} 
        & $P_r$ & 0.0003 & 0.0012 & 0.0008 & 0.0006 & 0.0018 & 0.0007 & 0.0010 & 0.0095 \\
        & $R_e$ & 0.7155 & 0.7850 & 0.7395 & 0.7104 & 0.7157 & 0.7785 & 0.7874 & 0.8069 \\
        & $F_1$ & 0.0005 & 0.0023 & 0.0015 & 0.0012 & 0.0035 & 0.0014 & 0.0019 & \textbf{0.0187} \\
        \bottomrule
    \end{tabular}
\end{table*}

Table \ref{ensemble_attention} presents a comparative evaluation of various ensemble learning models, including LightGBM, Random Forest (RF), XGBoost, Deep Forest (DF), AdaBoost, Extra Trees (ET), and Boosted Decision Trees (BDT), against the proposed attention-based stacking model for CCF detection. The models are assessed across three datasets and multiple evaluation metrics: $P_r$, $R_e$, and $F_1$. The proposed model consistently outperforms others, achieving the highest $P_r$, $R_e$, and $F_1$, especially in Dataset 1 ($F_1$ = 0.9992) and Dataset 2 ($F_1$ = 0.8556). Since the features in two datasets were anonymized, we assumed their equivalence when training on Dataset 1 and testing on Dataset 2. However, differences in underlying distributions due to potential variations in feature anonymization may have contributed to the lower $F_1$ score observed in Dataset 3 ($F_1$ = 0.0187). This suggests that while the proposed model generalizes well across datasets with similar feature representations, discrepancies introduced by anonymization could affect its ability to transfer knowledge effectively across different datasets.

Overall, the results across all three datasets confirm the effectiveness and adaptability of the proposed ensemble approach. On the balanced dataset (Dataset 1), the model demonstrates high reliability, particularly in correctly identifying legitimate transactions and maintaining a strong $R_e$ rate. When faced with a heavily imbalanced dataset (Dataset 2), the method preserves its comparative advantage over individual models, underscoring its robustness in challenging real-world conditions. Finally, in the cross-dataset evaluation (Dataset 3), the ensemble model again proves its versatility, striking an efficient trade-off between $R_e$ and $S_p$ when transitioning from a balanced training set to testing on imbalanced data.

\section{Discussion}
\label{discuss}
Fraud detection, which has gained significant research interest in recent years, has been tackled using a wide range of methodologies and algorithms. The following paragraphs highlight notable studies in this area.

Fanai et al. \cite{fanai2023novel} leveraged a deep Autoencoder for representation learning. The Autoencoder reduced the dimensionality of the data, extracting compact and informative representations, which were then used to train models, such as DNN, RNN, and a hybrid CNN-RNN. The Bayesian optimization algorithm was employed to fine-tune hyperparameters.

Tian et al. \cite{tian2023asa} employed adaptive sampling and aggregation-based graph neural network (ASA-GNN). This model addressed challenges such as noisy data, fraudsters’ camouflage behavior, and the over-smoothing issue in GNNs. ASA-GNN employed a neighbor sampling strategy that used cosine similarity to filter irrelevant nodes.

Zhu et al. \cite{zhu2023nus} proposed a noisy-sample-removed under-sampling scheme for imbalanced classification. This approach addressed the negative impact of noisy samples on classifier performance. The method utilized K-means clustering to remove noisy samples from both majority and minority classes.

The study \cite{cao2023feature} proposed an attention-based ensemble model. The model improved generalization and reduce variance by introducing a feature-wise attention mechanism and feature diversity regularization. The attention mechanism learned the most informative features, while the regularization term optimized the balance between bias and variance.

The study \cite{abdul2024federated} focus on a federated learning approach which enabled the development of a global model by aggregating locally computed updates from distributed datasets without sharing raw data, ensuring privacy. To tackle class imbalance, the study implemented both individual and hybrid resampling techniques, such as synthetic minority over-sampling technique (SMOTE) and adaptive synthetic (ADASYN).

Zhu et al. \cite{zhu2024adaptive} proposed an adaptive heterogeneous framework integrating deep RL and attention mechanisms. It used a reinforcement reward mechanism to select source domain instances and a CNN with an attention module to extract semantic features.

Tang et al. \cite{tang2024credit} concentrated on a federated graph learning approach, combining federated learning and GNN to model transaction relationships while preserving privacy. A graph extension algorithm and adaptive aggregation improved cross-institutional collaboration and addressed data imbalance.

\begin{table}[t]
\caption{Comparison summary of the proposed method with existing approaches.}
\label{disscusion_table}
\resizebox{\linewidth}{!}{%
\begin{tabular}{|l|l|l|l|l|l|l|}
\hline
Work &
  Dataset &
  Objective &
  Deep Learning Model &
  Performance \\ \hline
  2022\cite{xie2022time} &
  \begin{tabular}[c]{@{}l@{}} Public: 284 807 samples, \\ 492 fraudulent cases \\ \hline Private: 5 120 000 transactions, \\ 140 000 fraudulent records \\ \end{tabular} &
  \begin{tabular}[c]{@{}l@{}} Binary \\ \end{tabular} &
  \begin{tabular}[c]{@{}l@{}} Time-aware\\ historical-attention-\\ based LSTM \end{tabular} &
  \begin{tabular}[c]{@{}l@{}} $P_r$ = 0.504 \\ $R_e$ = 0.996 \\ $F_1$ = 0.669 \\ \hline $P_r$ = 0.921 \\ $R_e$ = 0.972 \\ $F_1$ = 0.946 \\ \end{tabular} \\ 
  \hline
  2023\cite{fanai2023novel} &
  \begin{tabular}[c]{@{}l@{}} Public: 284 807 samples, \\ 492 fraudulent cases \\ \hline Private: 1000 samples, \\ 30\% labled \\as bad credit \\ \end{tabular} &
  Binary &
  \begin{tabular}[c]{@{}l@{}} Deep neural network\\  + RNN\\ + Hybric CNN + RNN \end{tabular} &
  \begin{tabular}[c]{@{}l@{}} $F_1$ = 0.837 \\AUC = 0.921 \\ \hline $F_1$ = 0.807 \\AUC = 0.908  \end{tabular} \\ 
  \hline
  2023\cite{tian2023asa} &
  \begin{tabular}[c]{@{}l@{}} Private: 5 120 000 samples, \\ 140 000 fraudulent cases \\ \hline Public: 160 764 samples, \\ (28\% fraudulent cases) \\ \hline Public: 20 000 samples, \\ (balanced dataset) \\ \end{tabular} &
  Binary &
  \begin{tabular}[c]{@{}l@{}} Aggregation-based\\ graph neural \\ network \end{tabular} &
  \begin{tabular}[c]{@{}l@{}} $R_e$ = 0.884 \\ $F_1$ = 0.904 \\ AUC = 0.922 \\ \hline $R_e$ = 0.737 \\ $F_1$ = 0.774 \\ AUC = 0.835 \\ \hline $R_e$ = 0.731 \\ $F_1$ = 0.593 \\ AUC = 0.571 \\ \end{tabular} \\ 
  \hline
  2023\cite{zhu2023nus} &
  \begin{tabular}[c]{@{}l@{}} Private: 614 samples, \\ 192 fraudulent cases \\ \hline Private: 3 075 samples, \\ 448 fraudulent cases \\ \hline Private: 30 000 samples, \\ 6 636 fraudulent cases \\ \end{tabular} &
  Binary &
  \begin{tabular}[c]{@{}l@{}} Decision tree + \\ Logistic regression + \\ Random forest \\ as base classifier \\ in the NUS framework \end{tabular} &
  \begin{tabular}[c]{@{}l@{}} $F_1$ = 0.890 \\ G-mean = 0.910 \\ \hline $F_1$ = 0.860 \\ G-mean = 0.870 \\ \hline $F_1$ = 0.830 \\ G-mean = 0.850 \\ \end{tabular} \\ 
  \hline
  2023\cite{cao2023feature} &
  \begin{tabular}[c]{@{}l@{}} Public: 284 807 samples, \\ 492 fraudulent cases \\ \hline Private: 3 500 000 samples, \\ fraud rate 1.59\% \\ \end{tabular} &
  Binary &
  \begin{tabular}[c]{@{}l@{}} A feature-wise\\  attention-based\\ boosting ensemble\\ model \end{tabular} &
  \begin{tabular}[c]{@{}l@{}} $F_1$ = 0.885 \\ \hline $F_1$ = 0.929 \\ \end{tabular} \\ 
  \hline
  2024\cite{abdul2024federated} &
  \begin{tabular}[c]{@{}l@{}} Public: 284 807 samples, \\ 492 fraudulent cases \\ \end{tabular} &
  Binary &
  \begin{tabular}[c]{@{}l@{}} Federated CNN\\  models and \\ base classifiers \end{tabular} &
  \begin{tabular}[c]{@{}l@{}} $A_c$ = 0.999 \\ $P_r$ = 0.826 \\ $R_e$ = 0.809 \\ \end{tabular} \\ 
  \hline
  2024\cite{zhu2024adaptive} &
  \begin{tabular}[c]{@{}l@{}} Private: 5 125 107 samples, \\ 147 829 fraudulent cases \\ \hline Public: 284 807 samples, \\ 492 fraudulent cases \\ \end{tabular} &
  Binary &
  \begin{tabular}[c]{@{}l@{}} A CNN with\\  attention\\ mechanism \end{tabular} &
  \begin{tabular}[c]{@{}l@{}} $P_r$ = 0.921 \\ $F_1$ = 0.946 \\ \hline $P_r$ = 0.902 \\ $F_1$ = 0.880 \\ \end{tabular} \\ 
  \hline
  2024\cite{tang2024credit} &
  \begin{tabular}[c]{@{}l@{}} Public: 284 807 samples, \\ 492 fraudulent cases  \\ \hline 590 540 samples  \\ \end{tabular} &
  Binary &
  \begin{tabular}[c]{@{}l@{}} Federated graph\\  learning \end{tabular} &
  \begin{tabular}[c]{@{}l@{}} $P_r$ = 0.925 \\ $R_e$ = 0.910 \\ $F_1$ = 0.918 \\ $A_c$ = 0.935 \\ \hline $P_r$ = 0.710 \\ $R_e$ = 0.615 \\ $F_1$ = 0.640 \\ $A_c$ = 0.800 \\ \end{tabular} \\ 
  \hline
  2025\cite{yousefimehr2025distribution} &
  \begin{tabular}[c]{@{}l@{}} \\ Public: 284 807 samples, \\ 492 fraudulent cases \\ \end{tabular} &
  Binary &
  \begin{tabular}[c]{@{}l@{}} Sequential detection:\\  LSTM \\ \hline Non-Sequential \\ detection: \\ LightGBM \\ \end{tabular} &
  \begin{tabular}[c]{@{}l@{}}  $F_1$ = 0.850 \\ AUC = 0.870 \\ \hline $F_1$ = 0.870 \\ AUC = 0.960 \\ \end{tabular} \\ 
  \hline
  {Our work}&
  \begin{tabular}[c]{@{}l@{}} Public: 568 630 samples, \\ 284 615 fraudulent cases \\ \hline Public: 284 807 samples, \\ 492 fraudulent cases \\ \hline Public: 853 437 samples, \\ 284 807 fraudulent cases \\ \end{tabular} &
  Binary &
  \begin{tabular}[c]{@{}l@{}} Attention-based \\ ensemble of CNNs \\ and GNN\end{tabular} &
  \begin{tabular}[c]{@{}l@{}} $A_c$ = 0.9992\\ $F_1$ = 0.9992 \\ Pre = 0.9984\\ $R_e$ = 1.0 \\ \hline $A_c$= 0.9995 \\ $F_1$ = 0.8556 \\ Pre = 0.9091 \\ $R_e$ = 0.8081 \\ \hline $A_c$ = 0.830 \\ $R_e$ = 0.8069 \\ $S_p$ = 0.8539 \\ \end{tabular} \\ 
  \hline
  \end{tabular}%
}
\end{table}

The study \cite{yousefimehr2025distribution} developed integrated resampling methods and data-dependent classifiers. It combines one-class SVM with SMOTE and random under-sampling to preserve the distribution of fraud instances. The framework outputs are analyzed using LightGBM for non-sequential fraud detection and LSTM for sequential fraud detection. 

Table \ref{disscusion_table} compares this study’s approach with other recent fraud detection methods that employ deep neural networks. While many existing solutions have achieved respectable performance, our ensemble framework distinguishes itself through its uncertainty-aware design, combining DOWA, IOWA, and a confidence-driven gating mechanism, to effectively handle both balanced and highly imbalanced datasets. This architecture not only delivers strong $A_c$ and $R_e$ but also remains computationally feasible, making it adaptable for deployment in real-time payment gateways. Moreover, by integrating SHAP-based feature selection, each chosen attribute can be mapped to concrete transactional insights, thereby enhancing transparency for financial stakeholders. Such interpretability is pivotal for fostering trust in AI-driven fraud detection systems, as compliance teams and analysts can readily validate the rationale behind each flagged transaction.

\subsection{Prospects for Future Studies}
\label{recomm}
In future research, we aim to explore:
\begin{itemize}
    \item In complex tasks, by adding multiple layers of prediction, attention, and selection to the traditional stacking classifier structure, a richer model can identify complex patterns, which could be the focus of future research.
    \item Future studies are recommended to focus on other aggregation methods, such as fuzzy integral operators and approaches with a higher level of complexity than OWA operators.
    \item 
    Credit card transaction patterns can shift rapidly over time due to evolving fraud tactics or changes in consumer behavior. Incorporating online or incremental learning approaches could allow the model to adapt continuously and maintain high performance under these dynamic conditions.
    \item Deploying the proposed approach in real-world payment systems requires further examination of computational efficiency. Investigations into parallelization, cloud-based infrastructures, or lightweight model variants would help ensure the method can scale for high-throughput, real-time fraud detection scenarios.
    \item We randomly selected four models as the first layer classifiers. Future work could optimize the number of first layer classifiers by relying on forward selection and backward elimination approaches.

\end{itemize}

\section{Conclusion}
\label{Conclusion}
Given the accelerating shift toward digital transactions and the alarming surge in credit card fraud, effective detection systems have become critically important for financial security. 
This study introduced an attention-based stacking framework for CCF detection, incorporating both DOWA and IOWA aggregation strategies, along with a confidence-aware combination layer, to address the challenges of data imbalance, model uncertainty, and interpretability. Empirical evaluation on three datasets, ranging from balanced to heavily imbalanced, highlighted the method’s resilience and superiority over individual classifiers, consistently achieving strong $R_e$ and $S_p$. Additionally, by integrating SHAP to illuminate the most influential features, the model promotes transparency and fosters stakeholder trust. Notably, this framework is adaptable to varying computational constraints: in resource-limited settings, users may opt for one of the standalone classifiers to reduce complexity and speed up inference, whereas those with more robust computational resources can deploy the full ensemble approach for enhanced $A_c$ and robustness.

\bibliography{references}

\begin{thebibliography}{10}
\providecommand{\url}[1]{#1}
\csname url@samestyle\endcsname
\providecommand{\newblock}{\relax}
\providecommand{\bibinfo}[2]{#2}
\providecommand{\BIBentrySTDinterwordspacing}{\spaceskip=0pt\relax}
\providecommand{\BIBentryALTinterwordstretchfactor}{4}
\providecommand{\BIBentryALTinterwordspacing}{\spaceskip=\fontdimen2\font plus
\BIBentryALTinterwordstretchfactor\fontdimen3\font minus
  \fontdimen4\font\relax}
\providecommand{\BIBforeignlanguage}[2]{{%
\expandafter\ifx\csname l@#1\endcsname\relax
\typeout{** WARNING: IEEEtran.bst: No hyphenation pattern has been}%
\typeout{** loaded for the language `#1'. Using the pattern for}%
\typeout{** the default language instead.}%
\else
\language=\csname l@#1\endcsname
\fi
#2}}
\providecommand{\BIBdecl}{\relax}
\BIBdecl

\bibitem{mienye2024deep}
I.~D. Mienye and N.~Jere, ``Deep learning for credit card fraud detection: A
  review of algorithms, challenges, and solutions,'' \emph{IEEE Access}, 2024.

\bibitem{lebichot2024assessment}
B.~Lebichot, W.~Siblini, G.~M. Paldino, Y.-A. Le~Borgne, F.~Obl{\'e}, and
  G.~Bontempi, ``Assessment of catastrophic forgetting in continual credit card
  fraud detection,'' \emph{Expert Systems with Applications}, vol. 249, p.
  123445, 2024.

\bibitem{cherif2023credit}
A.~Cherif, A.~Badhib, H.~Ammar, S.~Alshehri, M.~Kalkatawi, and A.~Imine,
  ``Credit card fraud detection in the era of disruptive technologies: A
  systematic review,'' \emph{Journal of King Saud University-Computer and
  Information Sciences}, vol.~35, no.~1, pp. 145--174, 2023.

\bibitem{marchioni2023anomaly}
A.~Marchioni, A.~Enttsel, M.~Mangia, R.~Rovatti, and G.~Setti, ``Anomaly
  detection based on compressed data: An information theoretic
  characterization,'' \emph{IEEE Transactions on Systems, Man, and Cybernetics:
  Systems}, 2023.

\bibitem{madhurya2022exploratory}
M.~Madhurya, H.~Gururaj, B.~Soundarya, K.~Vidyashree, and A.~Rajendra,
  ``Exploratory analysis of credit card fraud detection using machine learning
  techniques,'' \emph{Global Transitions Proceedings}, vol.~3, no.~1, pp.
  31--37, 2022.

\bibitem{xie2022time}
Y.~Xie, G.~Liu, C.~Yan, C.~Jiang, and M.~Zhou, ``Time-aware attention-based
  gated network for credit card fraud detection by extracting transactional
  behaviors,'' \emph{IEEE Transactions on Computational Social Systems}, 2022.

\bibitem{gurun2018trust}
U.~G. Gurun, N.~Stoffman, and S.~E. Yonker, ``Trust busting: The effect of
  fraud on investor behavior,'' \emph{The Review of Financial Studies},
  vol.~31, no.~4, pp. 1341--1376, 2018.

\bibitem{shibata2022digitalization}
S.~Shibata, ``Digitalization or flexibilization? the changing role of
  technology in the political economy of japan,'' \emph{Review of International
  Political Economy}, vol.~29, no.~5, pp. 1549--1576, 2022.

\bibitem{barmo2024analysis}
A.~U. BARMO, A.~HARUNA, Y.~U. WALI, and K.~ABID, ``Analysis and comparison of
  fraud detection on credit card transactions using machine learning
  algorithms,'' 2024.

\bibitem{wang2021deep}
C.~Wang, Y.~Dou, M.~Chen, J.~Chen, Z.~Liu, and S.~Y. Philip, ``Deep fraud
  detection on non-attributed graph,'' in \emph{2021 IEEE International
  Conference on Big Data (Big Data)}.\hskip 1em plus 0.5em minus 0.4em\relax
  IEEE, 2021, pp. 5470--5473.

\bibitem{li2020deep}
Z.~Li, G.~Liu, and C.~Jiang, ``Deep representation learning with full center
  loss for credit card fraud detection,'' \emph{IEEE Transactions on
  Computational Social Systems}, vol.~7, no.~2, pp. 569--579, 2020.

\bibitem{khalid2024enhancing}
A.~R. Khalid, N.~Owoh, O.~Uthmani, M.~Ashawa, J.~Osamor, and J.~Adejoh,
  ``Enhancing credit card fraud detection: an ensemble machine learning
  approach,'' \emph{Big Data and Cognitive Computing}, vol.~8, no.~1, p.~6,
  2024.

\bibitem{han2022competition}
S.~Han, K.~Zhu, M.~Zhou, and X.~Cai, ``Competition-driven multimodal
  multiobjective optimization and its application to feature selection for
  credit card fraud detection,'' \emph{IEEE Transactions on Systems, Man, and
  Cybernetics: Systems}, vol.~52, no.~12, pp. 7845--7857, 2022.

\bibitem{zhao2022financial}
Z.~Zhao and T.~Bai, ``Financial fraud detection and prediction in listed
  companies using smote and machine learning algorithms,'' \emph{Entropy},
  vol.~24, no.~8, p. 1157, 2022.

\bibitem{ghosh2022spatio}
D.~K. Ghosh, A.~Chakrabarty, H.~Moon, and M.~J. Piran, ``A spatio-temporal
  graph convolutional network model for internet of medical things (iomt),''
  \emph{Sensors}, vol.~22, no.~21, p. 8438, 2022.

\bibitem{valavan2023predictive}
M.~Valavan and S.~Rita, ``Predictive-analysis-based machine learning model for
  fraud detection with boosting classifiers.'' \emph{Computer Systems Science
  \& Engineering}, vol.~45, no.~1, 2023.

\bibitem{chen2022credit}
M.~Chen, ``Credit card fraud detection based on multiple machine learning
  models,'' in \emph{Proceedings of the 2022 6th International Conference on
  Electronic Information Technology and Computer Engineering}, 2022, pp.
  1801--1805.

\bibitem{kowsalya2024credit}
K.~Kowsalya, M.~Vasumathi, and S.~Selvakani, ``Credit card fraud detection
  using machine learning algorithms,'' \emph{EPRA International Journal of
  Multidisciplinary Research (IJMR)}, vol.~10, no.~3, pp. 109--116, 2024.

\bibitem{chatterjee2024digital}
P.~Chatterjee, D.~Das, and D.~B. Rawat, ``Digital twin for credit card fraud
  detection: Opportunities, challenges, and fraud detection advancements,''
  \emph{Future Generation Computer Systems}, 2024.

\bibitem{abd2023efficient}
A.~Abd El-Naby, E.~E.-D. Hemdan, and A.~El-Sayed, ``An efficient fraud
  detection framework with credit card imbalanced data in financial services,''
  \emph{Multimedia Tools and Applications}, vol.~82, no.~3, pp. 4139--4160,
  2023.

\bibitem{chagahi2024cardiovascular}
M.~H. Chagahi, S.~M. Dashtaki, B.~Moshiri, and M.~J. Piran, ``Cardiovascular
  disease detection using a novel stack-based ensemble classifier with
  aggregation layer, dowa operator, and feature transformation,''
  \emph{Computers in Biology and Medicine}, p. 108345, 2024.

\bibitem{yager2016some}
R.~R. Yager and N.~Alajlan, ``Some issues on the owa aggregation with
  importance weighted arguments,'' \emph{Knowledge-Based Systems}, vol. 100,
  pp. 89--96, 2016.

\bibitem{dashtaki2022stock}
S.~M. Dashtaki, M.~Alizadeh, and B.~Moshiri, ``Stock market prediction using
  hard and soft data fusion,'' in \emph{2022 13th International Conference on
  Information and Knowledge Technology (IKT)}.\hskip 1em plus 0.5em minus
  0.4em\relax IEEE, 2022, pp. 1--7.

\bibitem{filev1998issue}
D.~Filev and R.~R. Yager, ``On the issue of obtaining owa operator weights,''
  \emph{Fuzzy sets and systems}, vol.~94, no.~2, pp. 157--169, 1998.

\bibitem{khaled2023dowg}
A.~Khaled, K.~Mishchenko, and C.~Jin, ``Dowg unleashed: An efficient universal
  parameter-free gradient descent method,'' \emph{Advances in Neural
  Information Processing Systems}, vol.~36, pp. 6748--6769, 2023.

\bibitem{wang2022study}
Y.~Wang, Y.~He, and Z.~Zhu, ``Study on fast speed fractional order gradient
  descent method and its application in neural networks,''
  \emph{Neurocomputing}, vol. 489, pp. 366--376, 2022.

\bibitem{krizhevsky2017imagenet}
A.~Krizhevsky, I.~Sutskever, and G.~E. Hinton, ``Imagenet classification with
  deep convolutional neural networks,'' \emph{Communications of the ACM},
  vol.~60, no.~6, pp. 84--90, 2017.

\bibitem{sherstinsky2020fundamentals}
A.~Sherstinsky, ``Fundamentals of recurrent neural network (rnn) and long
  short-term memory (lstm) network,'' \emph{Physica D: Nonlinear Phenomena},
  vol. 404, p. 132306, 2020.

\bibitem{zhou2020graph}
J.~Zhou, G.~Cui, S.~Hu, Z.~Zhang, C.~Yang, Z.~Liu, L.~Wang, C.~Li, and M.~Sun,
  ``Graph neural networks: A review of methods and applications,'' \emph{AI
  open}, vol.~1, pp. 57--81, 2020.

\bibitem{fanai2023novel}
H.~Fanai and H.~Abbasimehr, ``A novel combined approach based on deep
  autoencoder and deep classifiers for credit card fraud detection,''
  \emph{Expert Systems with Applications}, vol. 217, p. 119562, 2023.

\bibitem{tian2023asa}
Y.~Tian, G.~Liu, J.~Wang, and M.~Zhou, ``Asa-gnn: Adaptive sampling and
  aggregation-based graph neural network for transaction fraud detection,''
  \emph{IEEE Transactions on Computational Social Systems}, 2023.

\bibitem{zhu2023nus}
H.~Zhu, M.~Zhou, G.~Liu, Y.~Xie, S.~Liu, and C.~Guo, ``Nus:
  Noisy-sample-removed undersampling scheme for imbalanced classification and
  application to credit card fraud detection,'' \emph{IEEE Transactions on
  Computational Social Systems}, 2023.

\bibitem{cao2023feature}
R.~Cao, J.~Wang, M.~Mao, G.~Liu, and C.~Jiang, ``Feature-wise attention based
  boosting ensemble method for fraud detection,'' \emph{Engineering
  Applications of Artificial Intelligence}, vol. 126, p. 106975, 2023.

\bibitem{abdul2024federated}
M.~Abdul~Salam, K.~M. Fouad, D.~L. Elbably, and S.~M. Elsayed, ``Federated
  learning model for credit card fraud detection with data balancing
  techniques,'' \emph{Neural Computing and Applications}, vol.~36, no.~11, pp.
  6231--6256, 2024.

\bibitem{zhu2024adaptive}
K.~Zhu, N.~Zhang, W.~Ding, and C.~Jiang, ``An adaptive heterogeneous credit
  card fraud detection model based on deep reinforcement training subset
  selection,'' \emph{IEEE Transactions on Artificial Intelligence}, 2024.

\bibitem{tang2024credit}
Y.~Tang and Y.~Liang, ``Credit card fraud detection based on federated graph
  learning,'' \emph{Expert Systems with Applications}, p. 124979, 2024.

\bibitem{yousefimehr2025distribution}
B.~Yousefimehr and M.~Ghatee, ``A distribution-preserving method for resampling
  combined with lightgbm-lstm for sequence-wise fraud detection in credit card
  transactions,'' \emph{Expert Systems with Applications}, vol. 262, p. 125661,
  2025.

\end{thebibliography}
\end{document}